\documentstyle[epsf]{aipproc}
\newcommand{\BEQ}{\begin{equation}}
\newcommand{\EEQ}{\end{equation}}
\newcommand{\BEA}{\begin{eqnarray}}
\newcommand{\EEA}{\end{eqnarray}}

\renewcommand{\d}{{\rm d }}

\renewcommand{\S}{S_{\rm ep}}
\newcommand{\p}{\partial}

\newcommand{\I}{{\cal I}}

\def\dbarrm {{\mathchar'26\mkern-11mu{\rm d}}}

\newcommand{\fix}{{\Bigl.\Bigr|}}
\begin{document}
\title{Thermodynamics of the glassy state}

\author{Th.M. Nieuwenhuizen}
\address{Van der Waals-Zeeman Instituut,\\
Valckenierstraat 65, 1018 XE Amsterdam, The Netherlands}

\maketitle

\begin{abstract}

A picture for thermodynamics of the glassy state is introduced.
It assumes that one extra parameter, 
the effective temperature, is needed to
describe the glassy state. This explains the classical paradoxes
concerning the Ehrenfest relations and the Prigogine-Defay ratio.

As a second part, the approach connects the response of
macroscopic observables to a field change with
their temporal fluctuations, and with
the fluctuation-dissipation relation, in a generalized
non-equilibrium way. 

\end{abstract}

\section{Introduction}

Non-equilibrium thermodynamics for systems far from equilibrium
has long been a field of confusion. A typical application is window glass.
Such a system is far from equilibrium: a cubic micron of glass is neither
a crystal nor an ordinary under-cooled liquid.
It is an under-cooled liquid that, in the glass formation process,
 has fallen out of its meta-stable equilibrium. 

Until our recent works on this field, the general consensus reached
after more than half a century of research was:
{\it Thermodynamics does not work for glasses,
because there is no equilibrium}~\cite{Angell}. 
This conclusion was mainly based on
the failure to understand the Ehrenfest relations
and the  related Prigogine-Defay ratio. It should be kept in mind
that, so far, the approaches leaned very much on equilibrium ideas.
Well known examples are the 1951 Davies-Jones paper~\cite{DaviesJones},
 the 1958 Gibbs-DiMarzio
{}~\cite{GibbsDiMarzio} and the 1965 Adam-Gibbs~\cite{AdamGibbs} papers,
while a 1981 paper by DiMarzio has title ``Equilibrium theory of
glasses'' and a subtitle ``An equilibrium theory of glasses is
absolutely necessary''~\cite{DiMarzio1981}.
We shall stress that such approaches are not applicable, due to the
inherent non-equilibrium character of the glassy state.

Thermodynamics is the most robust field of physics.
Its failure to describe the glassy state is  quite unsatisfactory, 
since up to 25  decades in time can be involved.
Naively we expect that each decade has its own dynamics, 
basically independent of the other ones. 
We have found support for this point
in models that can be solved exactly.
Thermodynamics then means a description of system properties
under smooth enough non-equilibrium conditions.

\section{Thermodynamic picture for a system described by an
effective temperature} \label{General}

A state that slowly relaxes to equilibrium is characterized by the
time elapsed so far, sometimes called ``age'' or ``waiting time''.
For glassy systems this is of special relevance.
For experiments on spin glasses it is known that non-trivial cooling
or heating trajectories can be described by an effective age~\cite{Hammann}.
Yet we do not wish to discuss spin glasses. They have an
infinity of long time-scales, or infinite order replica symmetry
breaking. 

We shall restrict to systems with one diverging time scale,
having, in the mean field limit, one step of replica symmetry
breaking. They are systems with first-order-type phase transitions,
with discontinuous order parameter, though usually there is no latent heat.
(As we shall discuss, the same approach applies to true first order glassy
transitions that do have a latent heat.)

We shall consider glassy transitions for glass forming 
liquids as well as for random  magnets.
The results map onto each other by interchanging  volume $V$,
 pressure $p$, compressibility $\kappa=-\p \ln V/\p p$, and
expansivity $\alpha=\p \ln V/\p T$,
 by  magnetization $M$,  field $H$, susceptibility
$\chi=(1/N)\p M/\p H$,  and ``magnetizability'' $\alpha=(-1/N)\p M/\p T$,
respectively.

The picture to be investigated in this work starts
by describing a non-equilibrium state characterized by three parameters,
namely $T,p$ and the age $t$, or, equivalently, $T, p$ and the
{\it effective temperature} $T_e(t)$. This quantity has to follow
from solving the dynamics of the system, or from doing appropriate 
experiments.
For a set of smoothly related cooling experiments $T_i(t)$
at  pressures $p_i$, one may express the effective
temperature as a continuous function: $T_{e,i}(t)$ $\to$ $T_e(T,p)$.
This sets a surface in $(T,T_e,p)$ space, that becomes multi-valued
if one first cools, and then heats. For covering the whole space one
needs to do many experiments, e.g., at different pressures
and different cooling rates.
The results should agree with findings from heating experiments
and aging experiments.
Thermodynamics amounts to giving differential relations between
observables at nearby points in this space.

Of special importance is the thermodynamics of a thermal body at
temperature $T_2$ in a heat bath at temperature $T_1=T$. This could
apply to mundane situations such as a cup of coffee, or an ice-cream,
 in a room. There are also two entropies, $S_1$ and $S_2$. Notice that
there are also two time-scales: the time-scale for heat to leave the
cup is some ten minutes, while the time-scale for equilibrating that
heat in the room is much smaller. This separation of time-scales
allows the difference in temperatures. The change in heat of such
a system obeys $\dbarrm Q\le T_1\d S_1+T_2\d S_2$.

A similar two-temperature approach proves to be 
relevant for glassy systems.
The known exact results on the thermodynamics of systems without 
currents can be summarized by the very same change in heat
~\cite{NEhren} \cite{Nthermo}
\BEQ \label{dQ=}
\dbarrm Q=T\d\S+T_e\d\I
\EEQ
where $\S$ is the entropy of the fast or {\bf e}quilibrium
{\bf p}rocesses  ($\beta$-processes)
and $\I$ the configurational entropy of the slow or configurational
processes ($\alpha$-processes). This object is also known as
information entropy or complexity. 
In the standard definition~\cite{GibbsDiMarzio} 
the configurational entropy $\I$ is the entropy
of the glass minus the one of the vibrational modes of the crystal.
For polymers it still includes short-distance rearrangements, which
is a relatively fast mode. It was confirmed numerically
that $\I$ indeed does not vanish at any temperature,
thus violating the Adam-Gibbs relation between timescale and
configurational entropy~\cite{Binder}.
In our definition the configurational entropy
only involves long-time processes; the relatively
fast ones are counted in $S_{\rm ep}$. 
With this point of view the applicability of the 
Adam-Gibbs relation remains an open issue.

It is both surprising and
satisfactory that a glass can be described by the same general law.
If, in certain systems, also an effective pressure or field would be 
needed, then $\dbarrm Q$
is expected to keep the same form, but $\dbarrm W$ would change from its
standard value $-p\d V$ for liquids, or $-M\d H$ for magnets.
In the latter case it would become $-M_1\d H-M_2\d H_e$, where $H_e$
is the effective field, and $M_1$ and $M_2$ add up to $M$.
Such an extension could be needed to describe a larger class of systems.

\subsection{First and second law}

For a glass forming liquid the first law $\d U=\dbarrm Q+\dbarrm W$
becomes
\BEA \label{thermoglassp}
\label{dUp=}\d U=T\d \S+T_e \d \I-p\d V
\EEA
It is appropriate to define the generalized free enthalpy
\BEA \label{Gp=} G&=&U-T\S-T_e \I+pV\EEA
This is not the standard form, since $T_e\neq T$. It satisfies
\BEA
\label{dGp=}\d G&=&-\S\d T-\I\d T_e+V\d p
\EEA

The total entropy is
\BEQ \label{Stot=}S=\S+\I \EEQ
The second law requires $\dbarrm Q\le T\d S$, leading to
\BEQ (T_e-T)\d\I\le 0, \EEQ
which merely says that heat goes from high to low temperatures.

Since $T_e=T_e(T,p)$, and both entropies are functions of $T$, $T_e$
and $p$, the expression (\ref{dQ=}) yields a specific heat
\BEA
C_p&=&\frac{\p Q}{\p T}\fix_p=
T(\frac{\p \S}{\p T}\fix_{T_e,p}+
\frac{\p \S}{\p T_e}\fix_{T,p}\frac{\p T_e}{\p T}\fix_p)
+T_e(\frac{\p \I}{\p T}\fix_{T_e,p}+
\frac{\p \I}{\p T_e}\fix_{T,p}\frac{\p T_e}{\p T}\fix_p)
\EEA
In the glass transition region all factors, except $\p_T T_e$,
are basically constant. This leads to
\BEA \label{CpTool}
C_p&=&C_1+C_2\frac{\p T_e}{\p T}\fix_p
\EEA
Precisely this form has been assumed half a century ago
by Tool~\cite{Tool}  as starting point for
the study of caloric behavior in the glass formation region,
and has often been used for the explanation of experiments
{}~\cite{DaviesJones}\cite{Jaeckle86}.
It is  a direct consequence of eq. (\ref{dQ=}).

For magnetic systems the first law brings
\BEA
\label{dU=}\d U&=&T\d \S+T_e \d \I-M\d H
\EEA
As above, one can define the free energy
$F=U-T\S-T_e \I$. It satisfies the relation
$\d F=-\S\d T-\I\d T_e-M\d H $.

\subsection{Modified Maxwell relation}

For a smooth sequence of cooling procedures of a
glassy liquid, eq. (\ref{dUp=}) implies a modified
Maxwell relation between macroscopic observables
such as $U(t,p)\to U(T,p)= U(T,T_e(T,p),p)$ and $V$.
This solely occurs since $T_e$ is a non-trivial function of 
$T$ and $p$ for the smooth set of experiments under consideration. 

For glass forming liquids it reads
\BEQ \label{modMaxp}
\frac{\p U}{\p p}\fix_T + p\frac{ \p V}{\p p}\fix_T
+T\frac{\p V}{\p T}\fix_p
=T\frac{\p \I}{\p T}\fix_p\,\frac{\p T_e}{\p p}\fix_T-
T\frac{\p \I}{\p p}\fix_T\,\frac{\p T_e}{\p T}\fix_p+
T_e\frac{\p \I}{ \p p}\fix_T
\EEQ
This is the modified Maxwell relation between observables $U$ and $V$.
In equilibrium $T_e=T$, so the right hand side vanishes, and the
standard form is recovered.

Similarly, one finds for a glassy magnet
\BEQ \label{modMaxH}
\frac{\partial U}{\partial H}\fix_T+M-
T\frac{\partial M}{\partial T}\fix_H=
T_e\frac{\partial \I}{\partial H}\fix_T
+T\left(\frac{\partial T_e}{\partial H}\fix_T
\frac{\partial \I}{\partial T}\fix_H
-\frac{\partial T_e}{\partial T}\fix_H
\frac{\partial \I}{\partial H}\fix_T\right)
\EEQ

\subsection{Modified Clausius-Clapeyron relation}

Let us consider a first order transition between
two glassy phases A and B. An example could be the transition from
low-density-amorphous ice to high-density-amorphous ice 
~\cite{LDAHDAice}.
For the standard Clausius-Clapeyron relation
one uses that the free enthalpy $G$ is continuous along the
first order phase transition line $p_g(T)$. Since
$T_e\neq T$, it is actually not obvious that the function $G$
should still be continuous there. Based on a quasi-static approach 
involving a partition sum and, equivalently, on the experience
that in mean field models replica theory has always brought the
relevant physical free energy, we expect that our
generalized free enthalpy (\ref{Gp=}) is indeed continuous
at this transition.

Let us consider a first order transition between phases A and B.
Assuming that  $\Delta G =0$ along the glass transition line
$p_g(T)$, a little amount of algebra yields
the modified Clausius-Clapeyron relation
\BEQ
\Delta V \frac{\d p_g}{\d T} = \frac{\Delta U+p_g \Delta V}{T} +
\Delta\left(\frac{\d T_e}{ \d T}\I-\frac{T_e}{T}\I\right)
\EEQ
where $\d /\d T=\p/\p T+ (\d p_g/\d T)\p /\p p$ is the
``total'' derivative along the transition line.

It would be very interesting to test this relation for the
two forms of amorphous ice.
For those substances
Mishima and Stanley~\cite{HES} have presented a thermodynamic
construction of the standard free enthalpy or Gibbs potential $G$.
It is, however, based on equilibrium ideas and, 
in contrast to eq. (\ref{Gp=}), 
does not involve the effective temperature in the amorphous phases.
This approach therefore predicts the validity of 
the standard Clausius-Clapeyron relation. 
We feel that the standard $G$ is not the physically relevant one,
and that the analysis should be redone, by considering many cooling
rates and taking into account the
effective temperature and the possible violation of the 
Clausius-Clapeyron relation by going from the high density amorphous
phase to the low density one, and vice versa.

When phase A is an equilibrium under-cooled liquid, and phase B
is a glass, it holds that $T_e=T$ in phase A.
Then the relation reduces to
\BEQ
(V_A-V_B) \frac{\d p_g}{\d T} = \frac{U_A-U_B+p_g(V_A-V_B)}{T} +
(\frac{T_e}{T}-\frac{\d T_e}{ \d T})\I_B
\EEQ
Several models for magnets undergoing a first order glassy transitions
with a latent heat have been studied in the literature
~\cite{Mottishaw} \cite{Goldschmidt} \cite{ThirumDobrov}
\cite{NRitort}.

\subsection{Ehrenfest relations and Prigogine-Defay ratio}

In the glass transition region a glass forming liquid exhibits
smeared jumps in the specific heat $C_p$, the expansivity $\alpha$
and the compressibility $\kappa$. If one forgets about the smearing,
one may consider them as true discontinuities, yielding an analogy
with continuous phase transitions of the classical type.

Following Ehrenfest one may take the derivative of $\Delta
V(T,p_g(T))=0$. The result for a glass forming liquid may be written as
\BEQ \label{Ehren1p}
\Delta \alpha=\Delta \kappa \frac{\d p_g}{\d T}\EEQ
while for a glassy magnet
\BEQ \label{Ehren1H}
\Delta \alpha=\Delta \chi \frac{\d H_g}{\d T}\EEQ
The conclusion drawn from half a century of research on glass
forming liquids is that this relation is never satisfied
{}~\cite{DaviesJones}\cite{Goldstein}\cite{Jaeckle}~\cite{Angell}.
This has very much hindered progress on a thermodynamical approach.
However, from a theoretical viewpoint it is hard to imagine that
something could go wrong when just taking a derivative.
We have pointed out that this relation is indeed satisfied
automatically~\cite{NEhren}, but it is important say what is
meant by $\kappa$ in the glassy state.

Let us make an analogy with spin glasses. In mean field theory
they have infinite order replica symmetry breaking.
  From the early measurements of Canella and Mydosh ~\cite{Mydoshboek}
on AuFe it is known that
the susceptibility depends logarithmically on the frequency, so on
the time scale. The short-time value, called Zero-Field-Cooled (ZFC)
susceptibility is a lower bound, while the long time value, called
Field-Cooled (FC) susceptibility is an upper bound. Let us
use the term ``glassy magnets'' for  spin glasses
with one step of replica symmetry breaking. They are relevant for comparison
with glass forming liquids. For them the situation is worse, as the
ZFC value is discontinuous immediately below $T_g$.
 This explains why already directly below the glass transition
different measurements yield different values for $\kappa$.
These notions are displayed in figure 1.

\begin{figure}[b!] 
\label{chiplot}
\epsfxsize=10cm
\centerline{\epsffile{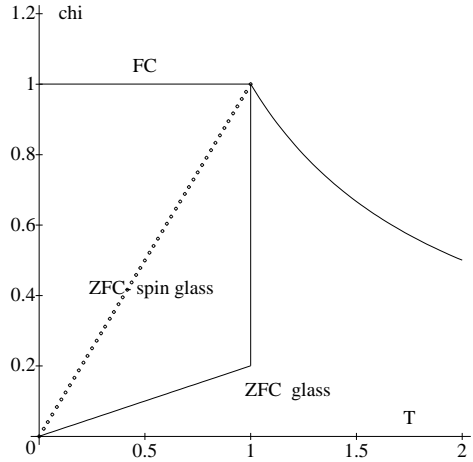}}
\caption{Schematic plot of the field-cooled (FC) and
zero-field-cooled (ZFC) susceptibility in
realistic spin glasses and  in glassy magnets,
as function of temperature, in arbitrary units.
In realistic spin glasses the infinite time or field-cooled
susceptibility is larger than the short time or zero-field-cooled
susceptibility. In magnetic analogs of realistic  glasses
the short time susceptibility even has a smeared  discontinuity at
the glass transition, yielding a value of $\chi$ that depends on 
the precise type of experiment which is performed. In glass forming
liquids the same happens for the compressibility. }
\end{figure}

 Previous claims about the violation of the first Ehrenfest relation
can be traced back to the equilibrium thermodynamics idea that there
 is one, ideal $\kappa$,  to be inserted in (\ref{Ehren1p}).
Indeed, investigators always considered cooling curves $V(T,p_i)$
at a set of pressures $p_i$ to determine $\Delta\alpha$ and
$\d p_g/\d T$. However, $\Delta \kappa$ was always determined in
another way, often from measurements of the speed of sound,
or by making more complicated pressure steps~\cite{RehageOels}.
In equilibrium such alternative determinations would yield the
same outcome. In glasses this is not the case: the speed of sound is
a short-time process, and additional pressure steps modify the glassy
state.  Therefore alternative procedures should be avoided, and only
the cooling curves  $V(T,p_i)$ should be used. They constitute
a liquid surface $V_{\rm liquid}(T,p)$ and a glass surface
$V_{\rm glass}(T,p)$ in $(T,p,V)$ space. These surfaces intersect,
and the first Ehrenfest relation is no more than a mathematical
identity about the intersection line of these surfaces.
It is therefore automatically satisfied~\cite{NEhren}.
The most careful data we came across were collected by Rehage and
Oels for atactic polystyrene\cite{RehageOels}. In figure 2
we present those data in a 3-d plot, underlining our point of view.
\begin{figure}[b!] 
\label{RehOelsplot3d}
\epsfxsize=15cm
\centerline{\epsffile{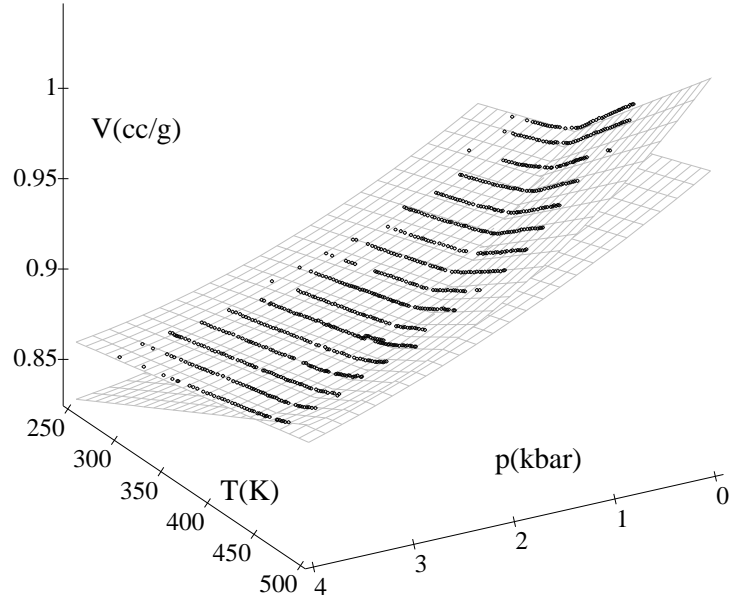}}
\caption{Data of the glass transition for cooling
atactic polystyrene at rate 18 $K/h$, scanned
  from the paper of Rehage and Oels (1976):
specific volume $V$ ($cm^3/g$) versus temperature $T$
($K$) at various pressures $p$ ($k\,bar$). As confirmed by a polynomial
fit, the data in the liquid
essentially lie on a smooth surface, and so do the data in the glass.
The first Ehrenfest relation describes no more than the
intersection of these surfaces, and is therefore
automatically satisfied. The values for the compressibility
derived in this manner will generally differ from results obtained
via other procedures.}
\end{figure}

The second Ehrenfest relation follows from differentiating $\Delta
U(T,p_g(T))=0$. The obtained relation will also be satisfied automatically.
However, one then eliminates $\partial U/\partial p$ by means of
the Maxwell relation. We have already discussed that
outside equilibrium it is modified.
As a result we obtain
\BEA \label{modEhren2p}
\frac{\Delta C_p}{T_gV}
&=&\Delta\alpha\frac{\d p_g}{\d T}
+\frac{1}{V}\left(1-\frac{\partial T_e}{\partial T}\fix_p\right)
\,\frac{\d \I}{\d T} 
\EEA
The $\I$ term constitutes the ``total'' derivative of the configurational 
entropy along the glass transition line.
Its prefactor only vanishes at equilibrium, in which case the
standard Ehrenfest relation is recovered.

For glassy magnets one has similarly
\BEQ \label{modEhren2H}
\frac{\Delta C}{NT}=\Delta \alpha\frac{\d H_g}{\d T}+
\frac{1}{N}\left(1-\frac{\partial T_e}{\partial T}\fix_H\right)
\,\frac{\d \I}{\d T}
\EEQ
The  equality
$T_e(T,p_g(T))=T$ implies the resulting identity
\BEQ \label{dTedT=1}
\frac{\d T_e}{\d T}=\frac{\p T_e}{\p T}\fix_p+
\frac{\p T_e}{\p p}\fix_T\frac{\d p_g}{\d T}=1 \EEQ
Combining the two Ehrenfest relations one may eliminate the
slope of the transition line. This leads to the so-called
Prigogine-Defay ratio
\BEQ
\Pi=\frac{\Delta C_p\Delta\kappa}{TV(\Delta \alpha)^2}
\EEQ
This looks like an equilibrium quantity.
For equilibrium transitions it should be
equal to unity. Assuming that at the glass transition a number of
unspecified parameters undergo a phase transition, Davies and Jones
derived that $\Pi\ge 1$~\cite{DaviesJones},
 while DiMarzio showed that in that case the correct value is $\Pi=1$
{}~\cite{DiMarzio}.
In glasses typical experimental values are reported in the range
$2<\Pi<5$. It was therefore generally expected that $\Pi\ge 1$ is
a strict inequality arising from the requirement of mechanical stability.

We have pointed out, however, that, since the first Ehrenfest relation is
satisfied, it holds that
\BEQ\label{Pip=}
\Pi=\frac{\Delta C_p}{T V\Delta\alpha (\d p_g/\d T)}=1+
\frac{1}{V\Delta \alpha }
\left(1-\frac{\partial T_e}{\partial T}\Bigl|_p \Bigr.\right)
\frac {\d \I}{\d p} \EEQ
Depending on the smooth set of experiments to be performed,
$\d p_g/\d T$ can be small or large: 
$\Pi$ {\it depends on the set of experiments}.
As a result, it can also be below
unity. Rehage-Oels found $\Pi=1.09\approx 1$ at $p=1$
$k\,bar$, using a short-time value for $\kappa$. Reanalyzing their data
we find from (\ref{Pip=}), where the correct $\kappa$ has been inserted,
 a value $\Pi=0.77$, which indeed is below unity.
The commonly accepted inequality $\Pi\ge 1$ is based on the
equilibrium assumption of a unique $\kappa$. 
Our theoretical arguments and the
Rehage-Oels data show that such idea's are incorrect.

\subsection{Fluctuation formula}

The basic result of statistical physics is that it relates
fluctuations in macroscopic variables to response of their averages
to changes in external field or temperature.
We have wondered whether such relations generalize to the glassy
state. We have found arguments in favor of such a possibility
both from the fluctuation-dissipation relation and by
exactly solving the dynamics of model systems ~\cite{Nhammer}.
Susceptibilities appear to have a non-trivial decomposition, that
looks as being very general. Here we give  arguments leading to it.

In cooling experiments at fixed field it holds that
$M=$$M(T(t),T_e(t,H),H)$. For thermodynamics one eliminates
time, implying
$M=$$M(T,T_e(T,H),H)$. One may then expect two terms:
\BEA\label{flucts=}
\chi&\equiv&\frac{1}{N}\,
\frac{\partial M}{\partial H}\Bigl|_T\Bigr.
= \chi^{\rm fluct}(t)+\chi^{\rm conf}(t)
\EEA
The first is the fluctuation contribution
\BEA \label{chifluct1}
\chi^{\rm fluct}(t)&=&\frac{1}{N} \frac{\partial M}{\partial H}
\Big|_{T,T_e}\Bigr. \EEA
To calculate it, we switch from a cooling experiment
to an aging experiment at the considered
$T$, $T_e$ and $H$, by keeping, in Gedanken, $T$ fixed from then on.
The system  will continue to age,  expressed by $T_e=T_e(t;T,H)$.
We may then use the equality
\BEQ\label{Mabhelp}
\frac{\partial M}{\partial H}\Big|_{T,t}\Bigr.
=\frac{\partial M}{\partial H}
\Big|_{T,T_e}\Bigr.+ \frac{\partial M}{\partial T_e}
\Big|_{T,H}\Bigr. \frac{\partial T_e}{\partial H}
\Big|_{T,t}\Bigr.
\EEQ
We have conjectured ~\cite{Nhammer}
that the left hand side may be written as the
sum of fluctuation terms for fast and slow processes,
\BEQ\label{Mabhelp1}
\frac{\partial M}{\partial H}\Big|_{T,t}\Bigr.
=\frac{ \langle \delta M^2(t)\rangle_{\rm fast}}{NT(t)}+
\frac{\langle \delta M^2(t)\rangle_{\rm slow}}{NT_e(t)}\EEQ
The first term is just the standard equilibrium expression for the fast
equilibrium processes.
Notice that the slow processes enter with their own temperature,
the effective temperature.  This decomposition
is confirmed by use of the fluctuation-dissipation relation
in the form to be discussed below.
Combination of (\ref{chifluct1}), (\ref{Mabhelp}) and
(\ref{Mabhelp1}) yields our non-equilibrium prediction
\BEA \label{chifluct}
\chi^{\rm fluct}(t)&=&
\frac{ \langle \delta M^2(t)\rangle_{\rm fast}}{NT(t)}+
\frac{\langle \delta M^2(t)\rangle_{\rm slow}}{NT_e(t)}
-\frac{1}{N}\frac{\partial M}{\partial T_e}\Big|_{T,H}\,\,
\frac{\p T_e}{\partial H}\Bigr|_{T,t}
\EEA
The first two terms are instantaneous, and thus the same for aging
and cooling. The third term is a correction, related to
an aging experiment. In the models  considered so far, 
this term is small~\cite{Nhammer}\cite{Nlongthermo}.

Since $T_e\neq T$, there occurs in eq. (\ref{flucts=})
also a new, configurational term
\BEA \label{chiconf}
\chi^{\rm conf}=\frac{1}{N}\,
\frac{\partial M}{\partial T_e}\Bigl|_{T,H}\,\,
\frac{\partial T_e}{\partial H}\Bigr|_{T}
\EEA
It originates from the difference in  the system's
structure for cooling experiments at nearby fields.
This is the term that is responsible for
 the discontinuity of $\chi$ at the glass transition.
For  glass forming liquids such a term occurs in the compressibility.
Its existence was anticipated in some earlier work~\cite{Goldstein} 
~\cite{Jaeckle}. 

\subsection{Fluctuation-dissipation relation}

Nowadays quite some attention is payed to the fluctuation-dissipation
relation in the aging regime of glassy systems. It was first put forward
in works by Horner~\cite{Horner1}\cite{CHS}
and generalized by Cugliandolo and Kurchan, 
see ~\cite{BCKM} for a review.

In the aging regime there holds a fluctuation-dissipation 
relation between the correlation function 
$C(t,t')$$=$$<\!\delta M(t)\delta M(t')\!>$ and $G(t,t')$, 
the response of $<\!M(t)\!>$ to a short, small field change 
$\delta H(t')$ applied at an earlier time $t'$,
\BEQ \label{FDR=}
\frac{\partial C(t,t')}{\p t'} = T_e(t,t'){G(t,t')}
\EEQ
with $T_e(t,t')$ being an effective temperature,
formally defined by this relation.
In the equilibrium or short-time regime $|t-t'|\ll t$, 
it is just equal to $T$; in the aging regime $t/t'={\cal O}(1)$, 
it depends only logarithmically on $t$ and $t'$, making it a useful
concept.

We have observed that in simple models without fast
processes $T_e(t,t')=\tilde T_e(t')$
is a function of one of the times only. 
One then expects that $\tilde T_e(t)$ is close to the
``thermodynamic'' effective temperature $T_e(t)$.
We have shown that ~\cite{Nlongthermo}
\BEQ \label{tildeTegen}
\tilde T_e(t)=T_e(t)-\dot T_e(t)
\left(\frac{ \p \ln C(t,t')}
{\p t'}\fix_{t'=t}\right)^{-1}+\cdots
\EEQ
So the effective temperatures $T_e$ and $\tilde T_e$
are not identical. However, in the models analyzed so far,
the difference is subleading in $1/\ln t$.

Notice that the ratio $\p_{t'}C(t,t')/G(t,t')=\tilde T_e(t')$
is allowed to depend on time $t'$.
The situation with constant $T_e$ is well known from mean
field spin glasses~\cite{BCKM}, but we have not  found such a
constant $T_e$ beyond mean-field~\cite{Nhammer}\cite{Nlongthermo}.
Only at exponential time-scales the mean field spin glass 
behaves as a realistic system~\cite{Nthermo}.

\newpage

\subsection{Time-scale arguments}

Consider a simple system that has only one type of processes ($\alpha$
processes), which falls out of equilibrium at some low $T$.
When it ages a time $t$ at $T=0$ it will have achieved a state
with effective temperature $T_e$, that can be estimated
by equating time with the equilibrium time-scale. Let us define
${\overline  T}_e$ by
\BEQ \label{barTe}
t=\tau_{eq}({\overline T}_e)\EEQ
We have checked in solvable models that,
to leading order in $\ln t$, it holds that ${\overline T}_e=T_e$.
(The first non-leading order turns out to be  non-universal).
This equality also
is found in cooling trajectories, when the system is well
inside the glassy regime. It says that the
system basically has forgotten its history, and ages on its own,
without caring about the actual temperature.
Another way of saying is that dynamics in each new
time-decade is basically independent of previous decade.

In less trivial systems, for instance those having a
Vogel-Tammann-Fulcher law,  the time-scale may have parameters
that depend on the actual temperature, implying $\tau=\tau(T,T_e)$.
We have already found support for the expectation that, 
to leading order, $T_e$ follows by equating this expression with time $t$.

In many systems one finds a $t'/t$ scaling
in the aging regime of two-time quantities.
There is a handwaving argument to explain that:
\BEQ
C(t,t')\approx C(\frac{t-t'}{\tau_{eq}(T_e(t'))})
\approx C(\frac{t-t'}{t'})=C(\frac{t}{t'})\EEQ
showing indeed the familiar $t/t'$ scaling. 
In the models studied so far we have found logarithmic  
scaling corrections ~\cite{Nhammer}\cite{Nlongthermo}.

\section{Solvable models}

The above relations have been tested in models of which the statics
is exactly solvable, and the dynamics is partially solvable,
such as the $p$-spin model~\cite{CHS}\cite{NEhren} and a directed
polymer model with glassy behavior ~\cite{Ndirpol}\cite{NEhren}.
Partial results follow from the backgammon model~\cite{FranzRitort},
\cite{GodrecheLuck}.
 
More interesting are  models with exactly solvable  
parallel Monte Carlo dynamics,
such as  independent harmonic oscillators~\cite{BPR}~\cite{Nhammer}
\cite{Nlongthermo} or independent spherical spins in a random field
~\cite{Nhammer}\cite{Nlongthermo}. In both cases a glassy
behavior occurs when cooling towards low temperatures.
A related model with a set of fast modes and a set of slow modes,
that have a Vogel-Fulcher-Tammann-Hesse law for the divergence
of the equilibrium time-scale, is currently under study.

The two-temperature approach put forward here also explains the
thermodynamics of black holes~\cite{Nblackhole} and
star clusters~\cite{Nstarcl}.

\end{document}